# An Approach to Complex Bayesian-optimal Approximate Message Passing

Gabor Hannak, Martin Mayer, Gerald Matz, Norbert Goertz
Institute of Telecommunications
Vienna University of Technology, Vienna, Austria
Email: {ghannak,mmayer,gmatz,ngoertz}@nt.tuwien.ac.at

*Abstract*—In this work we aim to solve the compressed sensing problem for the case of a complex unknown vector by utilizing the Bayesian-optimal structured signal approximate message passing (BOSSAMP) algorithm on the jointly sparse real and imaginary parts of the unknown. By introducing a latent activity variable, BOSSAMP separates the tasks of activity detection and value estimation to overcome the problem of detecting different supports in the real and imaginary parts. We complement the recovery algorithm by two novel support detection schemes that utilize the updated auxiliary variables of BOSSAMP. Simulations show the superiority of our proposed method against approximate message passing (AMP) and its Bayesian-optimal sibling (BAMP), both in mean squared error and support detection performance.

## I. INTRODUCTION

### A. Compressed Sensing

The theory of compressed sensing (CS) allows to solve underdetermined systems of equations with an additional sparsity constraint on the unknown signal vector. In particular, we consider the (estimation) problem

$$\mathbf{y} = \mathbf{A}\mathbf{x} + \mathbf{w}, \quad (1)$$

with the unknown $\mathbf{x} \in \mathbb{C}^N$ and known $\mathbf{A} \in \mathbb{R}^{M \times N}$ with $M < N$. The noisy measurement model involves a nonzero $\mathbf{w} \in \mathbb{C}^M$. When the sparsity level $K$ of $\mathbf{x}$ is known, i.e., that at most $K$ components are nonzero, one can recover $\mathbf{x}$ with a suitable *measurement matrix* $\mathbf{A}$. In particular, $\mathbf{A}$ has to fulfill the restricted isometry property (RIP) [1], for which a necessary condition on the number of measurements is $M > cK \log(N/K)$ with a constant $c$ independent of $K, N$. Literature is rich in solving this underdetermined equation on the reals both in the noisy and noiseless settings. Many greedy methods are based on iteratively finding the column(s) of $\mathbf{A}$ that, when scaled properly, best explain(s) the measurement $\mathbf{y}$, i.e., minimize(s) the iteratively reduced residual error (e.g. orthogonal matching pursuit) [2]. The $\ell_0$-regularized problem, i.e., minimzing the weighted sum of the $\ell_2$-error and the $\ell_0$-pseudonorm of the estimate, can be solved by e.g. iterative hard thresholding (IHT) [3]. Relaxing this nonconvex problem to the $\ell_1$-regularized problem leads to the convex least absolute shrinkage and selection operator (LASSO) [4], which can be solved by e.g. iterative soft thresholding (IST) [5] or convex optimization tools.

### B. The Probabilistic Approach

More recently, modelling the unknown as a random variable with a certain, sparsity enforcing prior (typically a Laplacian distribution i.i.d. over the entries) turned out to be beneficial. When $\mathbf{x}$ is interpreted as a realization of a random variable, one can use the probabilistic graphical model underlying the measurement equation [6] to derive approximate message passing schemes [7]. Moreover, the elaborated message passing schemes allow for a wider range of prior distributions for the unknown (e.g. dense or Bernoulli distributions), not only those featuring sparsity [8], [9]. These approximate message passing schemes turn out to be extremely efficient because they do not calculate actual messages individually but simplify to vector valued algorithms. In particular, they do not rely on matrix inversions, only on multiplications and additions, and converge after very few iterations, thus allow for very large $N$.

Many of the above algorithms require knowledge of the exact sparsity, which, in many applications, is not at hand and needs to be estimated first. Moreover, their generalization to the complex-valued problem ($\mathbf{x} \in \mathbb{C}^N$) is not straightforward.

Recently, Bayesian optimal structured signal approximate message passing (BOSSAMP) [12] was proposed for recovering vectors whose prior distributions and/or supports are identical, when acquired with the same measurement matrix. The algorithm's main idea is to assign likelihoods to the component estimates and exchange those between the vector estimate's individual components during iterations.

### C. Contribution

We generalize the Bayesian-optimal approximate message passing (BAMP) algorithm [8] in order to solve the complex-valued underdetermined linear system by separating the unknown vector into its the real and imaginary parts, and utilizing an algorithm that was designed to deal with group and joint sparsity. We specialize to the case of the complex Bernoulli-Gaussian prior distribution. Through numerical experiments we show that exploiting the strict joint sparsity of the real and complex parts is possible and beneficial for the recovery. Moreover, we formulate two novel support detection schemes that eliminate the need for the apriori knowledge of the sparsity.

This paper is organized as follows. In Section II we formulate the Bayesian approach to compressed sensing. In Section III the BAMP and the BOSSAMP algorithms are outlined and the complex approach is presented. In Section IV two support detection schemes are proposed. In Section V the numerical performance evaluation of the introduced methods is presented.

*Notation.* Uppercase (lowercase) boldface letters denote matrices (column vectors). For a matrix $\mathbf{A}$ (vector $\mathbf{a}$), $\mathbf{A}_s$ ($a_s$) denotes its $s$th column ($s$th entry), and $\mathbf{A}_\mathcal{S}$ ($\mathbf{a}_\mathcal{S}$) denotes the the matrix (vector) constituted by the columns (entries) of $\mathbf{A}$ ($\mathbf{a}$) that are indexed by the elements of the set $\mathcal{S}$, respectively. For a vector $\mathbf{a}$, $\mathrm{supp}(\mathbf{a})$ denotes the support of $\mathbf{a}$, that is, the set of indices of the nonzero entries, and $\|\mathbf{a}\|_0 = |\mathrm{supp}(\mathbf{a})|$ denotes the number of nonzero entries in $\mathbf{a}$. The identity matrix of dimension $M$ is denoted by $\mathbf{I}_M$. The Dirac delta function is $\delta(\cdot)$, and $(\mathcal{C})\mathcal{N}(\mathbf{x}; \boldsymbol{\mu}, \mathbf{C})$ denotes the value of the (complex) normal distribution pdf with mean $\boldsymbol{\mu}$ and covariance matrix $\mathbf{C}$ evaluated at $\mathbf{x}$.

## II. PROBLEM FORMULATION

### A. Introduction

Given the measurement model (1)

$$\mathbf{y} = \mathbf{A}\mathbf{x} + \mathbf{w}$$

and that $M < N$, direct "inversion" in order to solve for $\mathbf{x}$ is not possible. However, if $\mathbf{x}$ is treated as the realization of a random variable $\mathbf{x}$ with known pdf $p_\mathbf{x}(\mathbf{x})$, then $\mathbf{y}$ is also a random variable and the minimum mean squared error (MMSE) estimator of $\mathbf{x}$ is given by its conditional expectation given the measured vector $\mathbf{y}$:

$$\hat{\mathbf{x}} = \arg\min_{\tilde{\mathbf{x}}} \mathrm{E}_{\mathbf{x},\mathbf{w}}\left\{\|\mathbf{x} - \tilde{\mathbf{x}}\|_2^2 \,\big|\, \mathbf{y} = \mathbf{y}\right\}$$
$$= \mathrm{E}_{\mathbf{x},\mathbf{w}}\left\{\mathbf{x} \,|\, \mathbf{y} = \mathbf{y}\right\}. \quad (2)$$

To understand why solving the above is difficult, we write out the expectation (applying Bayes' rule)

$$\hat{\mathbf{x}} = \int_{\mathbb{R}^N} \mathbf{x} p_{\mathbf{x}|\mathbf{y}}(\mathbf{x}|\mathbf{y}) d\mathbf{x} = \frac{1}{p_\mathbf{y}(\mathbf{y})} \int_{\mathbb{R}^N} \mathbf{x} p_{\mathbf{y}|\mathbf{x}}(\mathbf{y}|\mathbf{x}) p_\mathbf{x}(\mathbf{x}) d\mathbf{x}. \quad (3)$$

Here, $p_{\mathbf{x}|\mathbf{y}}(\mathbf{x}|\mathbf{y})$ is the *posterior distribution*, i.e., the conditional probability density of $\mathbf{x}$ given a measured vector $\mathbf{y}$. If the noise $\mathbf{w}$ is Gaussian i.i.d., i.e., $\mathbf{w} \sim \mathcal{N}(0, \sigma_w^2 \mathbf{I}_M)$, we can write

$$p_{\mathbf{y}|\mathbf{x}}(\mathbf{y}|\mathbf{x}) = \prod_{m=1}^{M} \frac{1}{\sqrt{2\pi}\sigma_w} \exp\left(-\frac{\|y_m - (\mathbf{A}\mathbf{x})_m\|_2^2}{2\sigma_w^2}\right).$$

Further, if the signal components are also independent,

$$p_\mathbf{x}(\mathbf{x}) = \prod_{n=1}^{N} p_{\mathsf{x}_n}(x_n),$$

we can write (3) as

$$\hat{\mathbf{x}} = \frac{1}{p_\mathbf{y}(\mathbf{y})} \int_{\mathbb{R}^N} \mathbf{x} \prod_{m=1}^{M} \frac{1}{\sqrt{2\pi}\sigma_w} \exp\left(-\frac{\|y_m - (\mathbf{A}\mathbf{x})_m\|_2^2}{2\sigma_w^2}\right)$$
$$\prod_{n=1}^{N} p_{\mathsf{x}_n}(x_n) d\mathbf{x}. \quad (4)$$

Even though $\mathbf{x}$ can be decoupled into $N$ independent components, the first product in the integration requires the full vector $\mathbf{x}$, and the integration must be carried out over an $N$-dimensional space, which is not feasible in practice.

---

**Algorithm 1** BAMP

**Input:** $t = 0$, $\mathbf{z}^{(t)} = \mathbf{y}$, $\beta^{(t)} = \frac{1}{M}\|\mathbf{z}^{(t)}\|_2^2$
**do:**
1: $t \leftarrow t + 1$
2: $\mathbf{u}^{(t-1)} = \hat{\mathbf{x}}^{(t-1)} + \mathbf{A}^T \mathbf{z}^{(t-1)}$
3: $\beta^{(t-1)} = \frac{1}{M}\|\mathbf{z}^{(t-1)}\|_2^2$
4: $\hat{\mathbf{x}}^{(t)} = F\left(\mathbf{u}^{(t-1)}; \beta^{(t-1)}\right)$,
5: $\mathbf{z}^{(t)} = \mathbf{y} - \mathbf{A}\hat{\mathbf{x}}^{(t)} + \frac{1}{M}\mathbf{z}^{(t-1)}\sum_{n=1}^{N} F'\left(u_n^{(t-1)}; \beta^{(t-1)}\right)$
**while** $\|\mathbf{z}^{(t)} - \mathbf{z}^{(t-1)}\|_2^2/\|\mathbf{z}^{(t-1)}\|_2^2 > \epsilon_{TOL}$ **and** $t < t_{\max}$
**Output:** $\hat{\mathbf{x}} = \hat{\mathbf{x}}^{(t)}$, $\hat{\mathbf{u}} = \hat{\mathbf{u}}^{(t-1)}$, $\hat{\beta} = \beta^{(t-1)}$

---

## III. BAYESIAN-OPTIMAL APPROXIMATE MESSAGE PASSING

### A. The BAMP Algorithm

The BAMP algorithm, stated in [8], solves (4) approximately but efficiently for large $N$, given $\mathbf{y}$, $\mathbf{A}$ and the prior distribution $p_\mathbf{x}(\mathbf{x})$. The algorithm is stated in Algorithm 1 with the following functions involving the signal prior:

$$F(u_n; \beta) = \mathrm{E}_{\mathsf{x}_n}\left\{\mathsf{x}_n \,|\, \mathsf{u}_n = u_n\right\}, \quad (5)$$
$$F'(u_n; \beta) = \frac{\mathrm{d}}{\mathrm{d}u_n} F(u_n; \beta).$$

The conditional pdf leading to (5) is

$$f_{\mathsf{x}_n|\mathsf{u}_n}(x_n|u_n; \beta) = \frac{1}{\sqrt{2\pi\beta}} \exp\left(-\frac{1}{2\beta}(x_n - u_n)^2\right) \frac{p_{\mathsf{x}_n}(x_n)}{p_{\mathsf{u}_n}(u_n)},$$
$$n = 1, \ldots, N. \quad (6)$$

The variance $\beta$ of this distribution is computed in every iteration of BAMP (in the $t$th iteration $\beta^{(t)}$) and is strictly larger than the variance $\sigma_w^2$ of the $M$ noise components in the measurement model. Moreover, this pdf applies for $n = 1, \ldots, N$, whereas $\mathbf{y}$ has only dimension $M$. It has been proven (see e.g. [6], [10]) that asymptotically (as $N \to \infty$ and $M/N = \mathrm{const.}$) the pdf (6) represents a new *decoupled* measurement model

$$u_n = x_n + \tilde{w}_n, \quad n = 1, \ldots, N, \quad \text{with } \tilde{w}_n \sim \mathcal{N}(0, \beta). \quad (7)$$

Here, the *effective noise* $\tilde{w}_n$ combines the measurement noise $w_n$ and the *undersampling noise*, which results from the fact that $M < N$. Its Gaussian distribution results from the asymptoticity, i.e., that $N \gg 1$.

### B. Bernoulli-Gaussian Prior Implementation

Motivated by the practical significance (e.g. the complex Gaussian channel model prominently used in telecommunications [11]), we examine the *circularly-symmetric complex Bernoulli-Gaussian* prior distribution:

$$p_{\mathsf{x}_n}(x_n) = \gamma_n^{(0)} \delta(x_n) + (1 - \gamma_n^{(0)}) \mathcal{CN}(x_n; 0, \sigma_x^2), \quad (8)$$

with $\mathbf{x} = \mathbf{x}^{(R)} + j\mathbf{x}^{(I)}$ ($j = \sqrt{-1}$) and $\gamma_n^{(0)}$ the *prior zero probability* for component $n$. Because of the circular symmetry, i.e., $\forall k \neq l : \mathrm{E}\{\mathsf{x}_k \mathsf{x}_l\} = 0$, the real and imaginary

parts follow the distributions

$$p_{\mathsf{x}_n^{(R)}}(x_n^{(R)}) = \gamma_n^{(0)}\delta(x_n^{(R)}) + (1-\gamma_n^{(0)})\mathcal{N}(x_n^{(R)}; 0, \sigma_x^2/2), \quad (9)$$

$$p_{\mathsf{x}_n^{(I)}}(x_n^{(I)}) = \gamma_n^{(0)}\delta(x_n^{(I)}) + (1-\gamma_n^{(0)})\mathcal{N}(x_n^{(I)}; 0, \sigma_x^2/2), \quad (10)$$

respectively. The fact that $\mathbf{A} \in \mathbb{R}^{M \times N}$ allows to separate the real and imaginary parts of the measurement:

$$\mathbf{y}^{(R)} = \mathbf{A}\mathbf{x}^{(R)} + \mathbf{w}^{(R)} \quad (11)$$
$$\mathbf{y}^{(I)} = \mathbf{A}\mathbf{x}^{(I)} + \mathbf{w}^{(I)}. \quad (12)$$

Following a naive approach one can run two BAMP instances that estimate the two parts independently. The complex estimate $\hat{\mathbf{x}}$ is then simply the combination of the real and the imaginary estimates. In the following we will call this method the *complex BAMP (cBAMP)*.

The real and imaginary parts, however, are not completely independent: when looking at one component, it takes either the zero value both in its real and imaginary parts, or takes a value from the normal distribution both in its real and imaginary parts. We thus wish to utilize an algorithm that exploits this strict joint sparsity of $\mathbf{x}^{(R)}$ and $\mathbf{x}^{(I)}$. In particular, when a component $x_n$ has a real part estimate far from 0, the estimation of its imaginary part should utilize this information and vote stronger for a nonzero imaginary component, and vica versa. In order to formally treat this dependency, we define a latent random variable, the *activity variable* $\mathsf{a}_n$ with the Bernoulli distribution

$$p_{\mathsf{a}_n}(a_n) = \gamma^{(0)}\delta(a_n) + (1-\gamma^{(0)})\delta(1-a_n), \quad (13)$$

independently across the indices $n$. The realization of the activity variable, $a_n$, indicates whether $x_n$ is zero or a realization of the complex normal distribution. Then, we can formulate the prior pdf (8) conditioned on the activity variable as

$$p_{\mathsf{x}_n|\mathsf{a}_n}(x_n|a_n=0) = \delta(x_n) \quad (14)$$
$$p_{\mathsf{x}_n|\mathsf{a}_n}(x_n|a_n=1) = \mathcal{CN}(x_n; 0, \sigma_x^2) \quad (15)$$
$$= \mathcal{N}(x_n^{(R)}; 0, \sigma_x^2/2) \quad (16)$$
$$+ j\mathcal{N}(x_n^{(I)}; 0, \sigma_x^2/2). \quad (17)$$

This way, the two parts are connected via a mutual underlying vector $\mathbf{a} = (a_1, \ldots, a_N)^T$. Further, if the estimation of $\mathbf{a}$ and $\mathbf{x}$ are treated separately, the potential problem of acquiring nonidentical supports, as in the case of cBAMP, is resolved.

### C. BOSSAMP for Complex Signals

Recently, BOSSAMP [12] was proposed for solving estimation problems involving joint (and group) sparsity. Writing the measurements', the unknowns' and the noise's real and imaginary parts into matrices

$$\begin{aligned}\mathbf{Y} &= (\mathbf{y}^{(R)}, \mathbf{y}^{(I)}), \\ \mathbf{X} &= (\mathbf{x}^{(R)}, \mathbf{x}^{(I)}), \\ \mathbf{W} &= (\mathbf{w}^{(R)}, \mathbf{w}^{(I)}),\end{aligned} \quad (18)$$

the measurement equation becomes

$$\mathbf{Y} = \mathbf{A}\mathbf{X} + \mathbf{W}, \quad (19)$$

**Algorithm 2** BOSSAMP for complex signals

**Input:** $t = 0$, $\hat{\mathbf{x}}^{(\star),(t)} = \mathbf{0}_{N \times 1}$, $\mathbf{z}^{(\star),(t)} = \mathbf{y}^\star$, $\boldsymbol{\gamma}^{(\star),(t)} = (1-\gamma^{(0)})\mathbf{1}_{N \times 1}$, for $\star = R, I$

**do:**
1: $t = t + 1$
2: **for** $\star = R, I$:
3:   $\mathbf{u}^{(\star),(t-1)} = \hat{\mathbf{x}}^{(\star),(t-1)} + \mathbf{A}^T \mathbf{z}^{(\star),(t-1)}$
4:   $\beta^{(\star),(t-1)} = \frac{1}{M}\|\mathbf{z}^{(\star),(t-1)}\|_2^2$
5:   $\hat{\mathbf{x}}^{(\star),(t)} = F\left(\mathbf{u}^{(\star),(t-1)}; \beta^{(\star),(t-1)}, \boldsymbol{\gamma}^{(\star),(t-1)}\right)$
6:   $\mathbf{z}^{(\star),(t)} = \mathbf{y}^{(\star)} - \mathbf{A}\hat{\mathbf{x}}^{(\star),(t)}$
    $+ \frac{1}{M}\mathbf{z}^{(\star),(t-1)}\sum_{n=1}^{N} F'\left(u_n^{(\star),(t-1)}; \beta^{(\star),(t-1)}, \gamma_n^{(\star),(t-1)}\right)$
7: **for** $\star = R, I$ **do:**
8:   $\mathbf{l}^{(\star),(t)} = U(\mathbf{u}^{(\star),(t-1)}, \beta^{(\star),(t-1)}, \boldsymbol{\gamma}^{(\star),(0)})$
9:   $\boldsymbol{\gamma}^{(\star),(t)} = V(\mathbf{l}^{(\star),(t)})$
**while** $\sum_{\star = R,I} \|\mathbf{z}^{(\star),(t)} - \mathbf{z}^{(\star),(t-1)}\|_2^2 / \|\mathbf{z}^{(\star),(t-1)}\|_2^2 > \epsilon_{TOL}$ **and** $t < t_{\max}$
**Output:** $\hat{\mathbf{x}}^{(\star)} = \hat{\mathbf{x}}^{(\star),(t)}$, $\mathbf{u}^{(\star)} = \mathbf{u}^{(\star),(t-1)}$, $\beta^{(\star)} = \beta^{(\star),(t-1)}$ for $\star = R, I$

and one can directly apply BOSSAMP for jointly sparse vectors as described in Algorithm 2 to arrive at what we will call complex BOSSAMP (cBOSSAMP). As input, the a priori zero probability $P(\mathsf{a}_n = 0) = \gamma_n^{(0)}$ is necessary. For the complex Bernoulli-Gaussian prior, the *likelihood update* $U(\cdot, \cdot, \cdot)$ is defined (componentwise and omitting the iteration index) as

$$l_n^{(*)} = U(u_n^{(\star)}, \beta^{(\star)}, \gamma_n^{(0)}) =$$
$$\log \frac{\gamma_n^{(0)}}{1-\gamma_n^{(0)}} + \frac{1}{2}\left(\log\frac{\beta^{(\bar{\star})} + \sigma_x^2}{\beta^{(\bar{\star})}} - \frac{u_n^{(\star)^2}\sigma_x^2}{\beta^{(\bar{\star})}(\beta^{(\bar{\star})} + \sigma_x^2)}\right),$$
$$\star = R, I, \quad \bar{\star} = I, R, \quad (20)$$

which, in essence, extracts the novel likelihood information from the current real (imaginary) estimate and combines it with the likelihood of the imaginary (real) part (see [12]). The function $V(\cdot)$ transforms this into the updated prior probability as follows:

$$V(l_n^{(*)}) = \frac{1}{1 + \exp(-l_n^{(*)})}. \quad (21)$$

From the final estimates, $\hat{\mathbf{x}}^{(R)}$, $\hat{\mathbf{x}}^{(I)}$, the complex estimate $\hat{\mathbf{x}}$ can directly be read:

$$\hat{\mathbf{x}} = \hat{\mathbf{x}}^{(R)} + j\hat{\mathbf{x}}^{(I)}. \quad (22)$$

*Discussion.* Compared to the naive approach that runs two BAMPinstances independently, we expect the BOSSAMP based method to perform as follows: a) in the $M/N$ range where the BAMP on their own converge (at any sensible noise level), BOSSAMP will have no significant advantage; b) in a wide range of $M/N$ where (with arbitrary low additive noise) BAMP algorithm will not converge, BOSSAMP will because of the likelihood exchanges during iterations. These suppositions will be empirically validated in the numerical section.

## IV. SUPPORT DETECTION

After meeting the stopping criterion, similar to BAMP, which delivers a MMSE estimate, the estimate $\hat{\mathbf{x}}$ rarely will have exact zero components. However, generally in CS, when the prior nonzero probabilities are equal ($\gamma_n^{(0)} = \gamma^{(0)} \forall n$), approximately $\gamma^{(0)} N$ components of the unknown vector are exactly zero. In many applications, one is interested in the support of the unknown, i.e., the indices of the (non)zero components. Therefore, after converging and acquiring the BOSSAMP estimate, we wish to detect the true nonzero components. To accomplish this goal, we call the decoupled measurement model (7) and the updated prior probabilities (of the last iteration) $\boldsymbol{\gamma}^{(\star)} = \boldsymbol{\gamma}^{(\star),(t)}$ into action.

### A. Prior-based Support Detection

The prior-based support detection follows the rule

$$\hat{x}_n \leftarrow 0 \quad \text{if} \quad \prod_{\star=R,I} \gamma_n^{(\star)} \geq \prod_{\star=R,I} (1-\gamma_n^{(\star)}), \quad n=1,\ldots,N. \quad (23)$$

Note that this rule does not use any amplitude information, but is directly applicable and computationally cheap.

### B. EM-based Support Detection

It is clear that if $\hat{u}_n$ has a large magnitude, based on the decoupled measurement model one can be confident that $x_n \neq 0$. On the other hand, if $\hat{u}_n$ is close to zero, we cannot be sure whether $\hat{x}_n$ is a noisy estimate of $x_n = 0$ or a (noisy) estimate of a small but nonzero $x_n$. As suggested by (7), the entries of both $\mathbf{u}^{(R)} = \mathbf{u}^{(R),(t-1)}$ and $= \mathbf{u}^{(I)} = \mathbf{u}^{(I),(t-1)}$ stem from one of two Gaussian distributions:

$$u_n^{(\star)} \sim \begin{cases} \mathcal{N}(0, \beta^{(\star)}) & \text{for } n \notin \text{supp}(\mathbf{x}), \\ \mathcal{N}(0, (\sigma_x^2/2 + \beta^{(\star)})) & \text{for } n \in \text{supp}(\mathbf{x}). \end{cases}$$

A probabilistically sound way of (soft) clustering vectors (numbers) that are assumed to come from different distributions is the expectation-maximization (EM) algorithm [13]. (For Gaussian distributions,) the EM algorithm not only classifies the vectors (*E-step*), but also finds its parameters (mean, (co-)variance) in the *M-step*. In our case, these parameters are already known, thus only a single E-step is necessary for classification. The E-step calculates the so called responsibilities. With the shorthand notation $\bar{\beta}^{(\star)} = \beta^{(\star)} + \sigma_x^2/2$:

$$\sigma_{00} = \gamma_n^{(R)} \gamma_n^{(I)} \mathcal{N}(u_n^{(R)}; 0, \beta^{(R)}) \mathcal{N}(u_n^{(I)}; 0, \beta^{(I)}),$$
$$\sigma_{01} = \gamma_n^{(R)} (1-\gamma_n^{(I)}) \mathcal{N}(u_n^{(R)}; 0, \beta^{(R)}) \mathcal{N}(u_n^{(I)}; 0, \bar{\beta}^{(I)}),$$
$$\sigma_{10} = (1-\gamma_n^{(R)}) \gamma_n^{(I)} \mathcal{N}(u_n^{(R)}; 0, \bar{\beta}^{(R)}) \mathcal{N}(u_n^{(I)}; 0, \beta^{(I)}),$$
$$\sigma_{11} = (1-\gamma_n^{(R)})(1-\gamma_n^{(I)}) \mathcal{N}(u_n^{(R)}; 0, \bar{\beta}^{(R)}) \mathcal{N}(u_n^{(I)}; 0, \bar{\beta}^{(I)}),$$

the responsibilities for the $n$th component being explained by the effective noise or the nonzero signal plus effective noise, respectively, are

$$\rho(\mathsf{a}_n = 0) \propto P\left(\mathsf{a}_n = 0 \mid (\mathsf{u}_n^{(R)}, \mathsf{u}_n^{(I)}) = (u_n^{(R)}, u_n^{(I)})\right)$$
$$\propto \frac{\sigma_{00}}{\sum_{i=0}^{1} \sum_{j=0}^{1} \sigma_{ij}}, \quad (24)$$

$$\rho(\mathsf{a}_n = 1) \propto P\left(\mathsf{a}_n = 1 \mid (\mathsf{u}_n^{(R)}, \mathsf{u}_n^{(I)}) = (u_n^{(R)}, u_n^{(I)})\right)$$
$$\propto \frac{\sigma_{11}}{\sum_{i=0}^{1} \sum_{j=0}^{1} \sigma_{ij}}. \quad (25)$$

The responsibility is the "relative contribution" of a particular distribution to the observation $u_n$. Using the fact that the denominators of $\rho(\cdot)$ are identical, transforming this soft classification into a hard clustering based on the responsibility values is straightforward and numerically efficient:

$$\hat{x}_n \leftarrow 0 \text{ if } \sigma_{00} \geq \sigma_{11}. \quad (26)$$

This method is computationally much more challanging to evaluate when $N$ gets large.

In the case of cBAMP the vectors $\mathbf{u}^{(\star)}$ are available, whereas no updated $\boldsymbol{\gamma}^{(\star)}$ values are acquired. Thus, the EM-based support detection for cBAMP utilizes the prior nonzero probabilities $\boldsymbol{\gamma}^{(0)}$ in (24)-(25).

After applying one of the two schemes, the detected support is $\text{supp}(\hat{\mathbf{x}})$.

## V. NUMERICAL RESULTS

In order to empirically evaluate the performances, we compare three aspects of complex BOSSAMP and other recovery methods (BAMP and approximate message passing (AMP)). First, we demonstrate by means of the *empirical phase transition curves* of the three algorithms that complex BOSSAMP allows for a lower undersampling ratio, i.e., lower $M/N$ at constant $K/N$, respective for lower sparsity, i.e., higher $K/N$ at constant $M/N$. Secondly, we compare the normalized mean squared error (NMSE) behaviour over varying $M$ and signal-to-noise ratio (SNR). Thirdly, we compare the support detection performances of the algorithms.

In all simulations $N = 1000$ was used, the maximum number of allowed iterations is $t_{\max} = 100$ and $\epsilon_{TOL} = 10^{-4}$. The entries of the measurement matrix $\mathbf{A}$ are uniform i.i.d. Bernoulli distributed with $A_{m,n} \in \{-1/\sqrt{M}, 1/\sqrt{M}\}$ such that the columns are normalized.

The phase transition curve for the noiseless case ($\mathbf{w} = 0$) is the set of points in the $M/N - K/M$ unit square where the probability of a *successful recovery* is 0.5. This curve separates the two halves of the unit square with parameters that allow for successful and unsuccessful recovery. We call a recovery successful recovery when the NMSE defined as

$$\text{NMSE} = \|\hat{\mathbf{x}} - \mathbf{x}\|_2^2 / \|\mathbf{x}\|_2^2 \quad (27)$$

is below $\epsilon_{TOL}$. In order to acquire the empirical phase transitions, we produce 200 independent random realizations of $\mathbf{A}, \mathbf{x}$ on every point on a $19 \times 19$ uniform grid in the $M/N - K/M$ square $[0.05, 0.95] \times [0.05, 0.95]$. For demonstrational purposes, we connect the points representing parameters with 50% success ratio to approximate the phase transition curves. In Fig. 1 we can observe the expected superior performance of cBOSSAMP relative to cBAMP and AMP (implemented according to [7] with the MSE minimizing heuristic $\lambda = 2.678K^{-0.181}$ [14]).

Similar to the phase transition curves we are interested in the isolines representing the parameters which lead to a *successful support detection* with probability 0.5. A support detection is successful if $\text{supp}(\mathbf{x}) = \text{supp}(\hat{\mathbf{x}})$, i.e., there are neither false negatives (missed detections) nor false positives (false alarms) in the detected support. Fig. 2 shows the empirical support detection phase transition curves in the undersampling ($M/N$) regime indicative of CS, generated similar to Fig. 1. It can clearly be observed that cBOSSAMP with both support detection schemes performs superior to cBAMP with the EM based support detection scheme. Furthermore, this simulation demonstrates that although the EM-based method uses amplitude information and is numerically much more challenging, there is no significant advantage over the simple scheme using only the $\gamma^{(\star)}$ values.

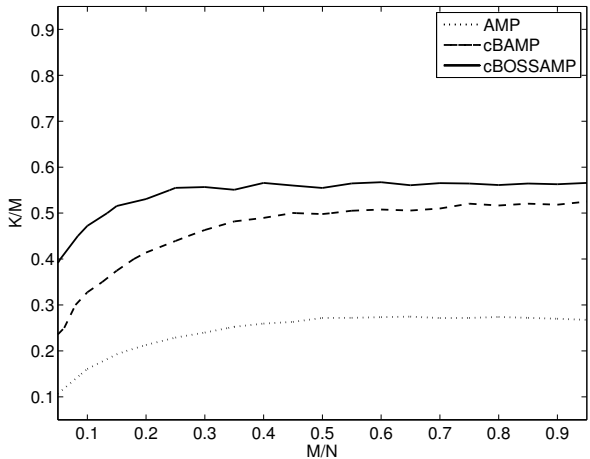

Fig. 1. Empirical phase transitions of AMP, cBAMP and cBOSSAMP.

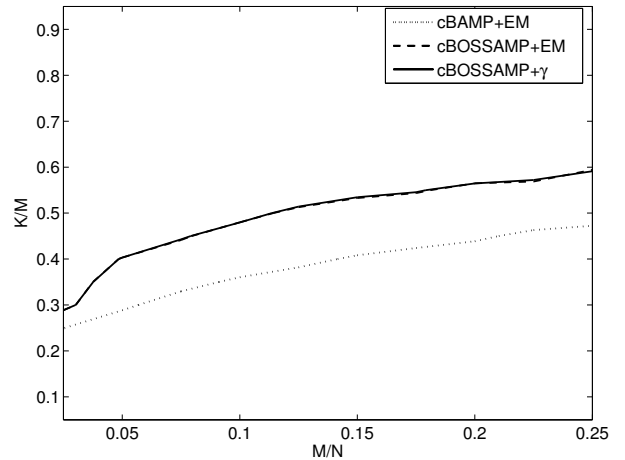

Fig. 2. Empirical support detection phase transitions of cBAMP and cBOSSAMP with the EM detection rule (EM) and cBOSSAMP with the prior-based detection rule ($\gamma$).

Fig. 3 shows the recovery NMSE results for sparsity $K = 20$, and number of measurements $M = 70$ and $M = 140$, respectively, for the three recovery methods AMP, cBAMP and the proposed cBOSSAMP. The SNR is defined as

$$\text{SNR} = \text{E}_{\mathbf{x},\mathbf{w}} \left\{ \frac{\|\mathbf{A}\mathbf{x}\|_2^2}{\|\mathbf{w}\|_2^2} \right\}, \tag{28}$$

and we averaged the results over 1000 indepentent realizations (at each SNR point) of $\mathbf{A}$, $\mathbf{x}$, $\mathbf{w}$. Observe that at sufficiently large $M$, cBAMP approaches the cBOSSAMP performance with growing SNR, whereas for lower $M$, cBAMP has a significant error floor and cBOSSAMP with only half as many measurements even outperforms AMP.

## VI. Conclusion and Outlook

We have applied the BOSSAMP algorithm successfully to recover complex vectors of high dimension in an underdetermined measurement setup. As an example, we implemented the complex Bernoulli-Gaussian prior. We have also proposed two novel support detection schemes that are applicable within the Bayesian framework. Numerical experiments have shown that our proposed method outperforms the naive generalizations of the state of the art algorithms to complex signals, both in NMSE performance and support detection capability. The implementation to other useful priors is straightforward and has potential in many applications.

## Appendix

The computation of the function

$$F(u;\beta) = \text{E}_{\mathsf{x}} \{\mathsf{x} \,|\, \mathsf{u} = u\} \tag{29}$$

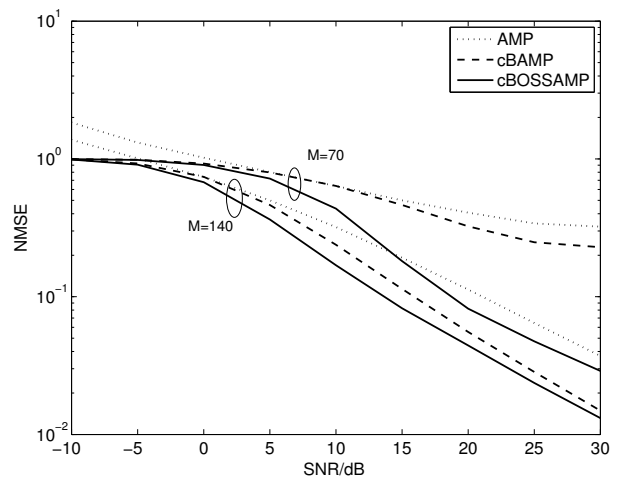

Fig. 3. NMSE behavior over the SNR of AMP, cBAMP and cBOSSAMP with $K = 20$ and $M = 70, 140$.

is of rather technical nature and thus we omit the presentation of the exact steps of calculation. It can be expressed as

$$F(u;\beta) = u - \sqrt{\frac{2\beta}{\pi}} \frac{I(u;\beta)}{J(u;\beta)} \tag{30}$$

with

$$\begin{aligned} I(u;\beta) &= \int_{-\infty}^{\infty} \frac{dp_{\mathsf{x}}(x)}{dx} \exp\left(-\frac{(x-u)^2}{2\beta}\right) dx \\ J(u;\beta) &= \int_{-\infty}^{\infty} \frac{dp_{\mathsf{x}}(x)}{dx} \text{erf}\left(-\frac{x-u}{\sqrt{2\beta}}\right) dx \,, \end{aligned} \tag{31}$$

with $\text{erf}(\cdot)$ being the Gauss error function. For the real Bernoulli-Gaussian prior

$$p_{\mathsf{x}}(x) = \gamma \delta(x) + (1-\gamma)\delta(1-x) \tag{32}$$

these integrals can be determined analytically by standard tools.

ACKNOWLEDGEMENTS

Part of this work was supported by the EU project NEWCOM# (GA 318306).